\documentclass[conference]{IEEEtran}
\IEEEoverridecommandlockouts

\usepackage{cite}

\usepackage[utf8]{inputenc}
\usepackage{algorithm}
\usepackage{algorithmic}

\usepackage{bm}
\usepackage{graphicx}
\usepackage{subfigure}
\usepackage{amsmath}
\usepackage{CJKutf8}
\usepackage{multirow}
\usepackage{textcomp}
\usepackage{balance}
\usepackage{tablefootnote}
\usepackage{footnote}

\newcommand{\thetaf}{\bm{\theta}_f}
\newcommand{\thetab}{\bm{\theta}_b}
\newcommand{\btheta}{\bm{\theta}}
\newcommand{\x}{\mathbf{x}}
\newcommand{\y}{\mathbf{y}}
\newcommand{\D}{\mathcal{D}}
\newcommand{\Y}{\mathcal{Y}}

\newcommand{\eg}{\textit{e.g.}}
\newcommand{\ie}{\textit{i.e.}}

\newcommand\blfootnote[1]{%
  \begingroup
  \renewcommand\thefootnote{}\footnote{#1}%
  \addtocounter{footnote}{-1}%
  \endgroup
}
\makeatletter
    \def\footnoterule{\kern-3\p@
      \hrule \@width 2in \kern 2.6\p@} 
    \makeatother

\DeclareRobustCommand*{\IEEEauthorrefmark}[1]{%
  \raisebox{0pt}[0pt][0pt]{\textsuperscript{\footnotesize #1}}%
}
\makeatletter
 \let\old@ps@headings\ps@headings
 \let\old@ps@IEEEtitlepagestyle\ps@IEEEtitlepagestyle
 \def\confheader#1{%
 \def\ps@headings{%
 \old@ps@headings%
 \def\@oddhead{\strut\hfill#1\hfill\strut}%
 \def\@evenhead{\strut\hfill#1\hfill\strut}%
 }%
 \def\ps@IEEEtitlepagestyle{%
 \old@ps@IEEEtitlepagestyle%
 \def\@oddhead{\strut\hfill#1\hfill\strut}%
 \def\@evenhead{\strut\hfill#1\hfill\strut}%
 }%
 \ps@headings%
 }
\makeatother

\confheader{%
Published as a conference paper at ICDE 2021}

\begin{document}

\title{Query Rewriting via Cycle-Consistent Translation for E-Commerce Search}

\author{\IEEEauthorblockN{Yiming Qiu\IEEEauthorrefmark{1}$^\dagger$,
Kang Zhang\IEEEauthorrefmark{1}$^\dagger$,
Han Zhang\IEEEauthorrefmark{1}, 
Songlin Wang\IEEEauthorrefmark{1},
Sulong Xu\IEEEauthorrefmark{1},
Yun Xiao\IEEEauthorrefmark{2},
Bo Long\IEEEauthorrefmark{1},
Wen-Yun Yang\IEEEauthorrefmark{2}$^*$}

\IEEEauthorblockA{\IEEEauthorrefmark{1}JD.com, Beijing, China}
\IEEEauthorblockA{\IEEEauthorrefmark{2}JD.com Silicon Valley Research Center, Mountain View, CA, United States}}

\maketitle
\begin{abstract}
Nowadays e-commerce search has become an integral part of many people's shopping routines. One critical challenge in today's e-commerce search is the semantic matching problem where the relevant items may not contain the exact terms in the user query. In this paper, we propose a novel deep neural network based approach to query rewriting, in order to tackle this problem. Specifically, we formulate query rewriting into a cyclic machine translation problem to leverage abundant click log data. Then we introduce a novel cyclic consistent training algorithm in conjunction with state-of-the-art machine translation models to achieve the optimal performance in terms of query rewriting accuracy. In order to make it practical in industrial scenarios, we optimize the syntax tree construction to reduce computational cost and online serving latency. Offline experiments show that the proposed method is able to rewrite hard user queries into more standard queries that are more appropriate for the inverted index to retrieve. Comparing with human curated rule-based method, the proposed model significantly improves query rewriting diversity while maintaining good relevancy. Online A/B experiments show that it improves core e-commerce business metrics significantly. Since the summer of 2020, the proposed model has been launched into our search engine production, serving hundreds of millions of users.

\end{abstract}


\blfootnote{$^\dagger\,$ Both authors contributed equally}
\blfootnote{$^*\,$ Corresponding author at wenyun.yang@jd.com}

\section{Introduction}

Over recent decades, online shopping platforms (e.g., eBay, Walmart, Amazon, Tmall, Taobao and JD) have become increasingly popular in people's daily life. E-commerce search, which helps users to find what they need from billions of products, is an essential part of those platforms, contributing to the largest percentage of transactions among all channels. 
For instance, the top e-commerce platforms in China, \eg, Tmall, Taobao and JD, serve hundreds of million active users with gross merchandise volume of hundreds of billion US dollars.

However, a lot of e-commerce search queries do not have satisfactory results from traditional search engines. This is due to the nature of e-commerce search: a) item titles are often short, thus hard for the inverted index to retrieve, b) significant numbers of new internet users tend to make natural language alike search queries, \eg, ``cellphone for grandpa'', ``gift for girlfriend'', c) polysemous queries are more common in e-commerce search, \eg, ``apple'' could mean Apple company's products or the fruit apple.
Based on our internal analysis of real log data, these three cases causes most of the unsatisfactory search results in our company's search engine, one of the largest e-commerce search platform in the world. 

The traditional search engine typically performs the candidate retrieval stage using an inverted index, which is built to efficiently retrieve candidate items based on term matching. This stage greatly reduces the number of candidates from billions to thousands. It is a core step in the search engine.
However, due to the term mismatch between query and item titles, the candidate retrieval stage contributes most to the failure cases in our production.

\begin{figure}[t]
    \centering
    \includegraphics[width=80mm]{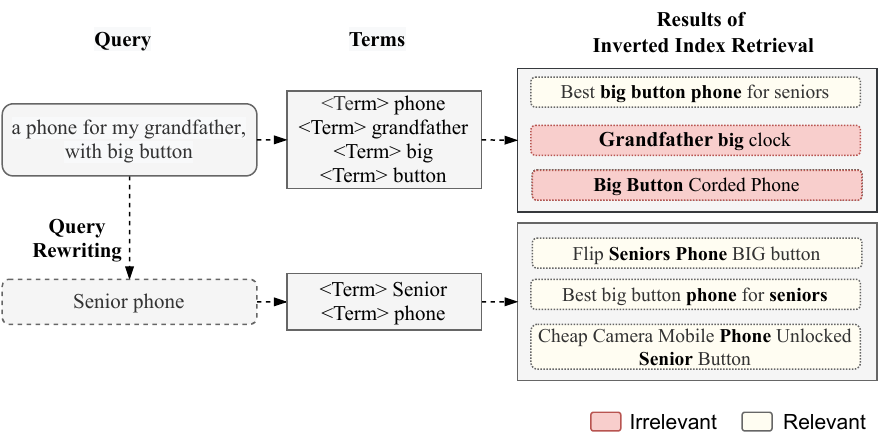}
    \caption{Illustration of query rewriting process that retrieves more relevant results.}
    \label{fig:example}
    \vspace{-3mm}
\end{figure}

In academia, \emph{semantic matching} refers to this kind of term mismatch problems, \ie, semantically relevant items cannot be retrieved by inverted indexes, since the item title does not contain the exact terms of a query. This is particularly common in e-commerce search because item titles are often short. 
We notice that a lot of long-tail queries or natural language alike queries are failed in this way. As an example, it is almost impossible to retrieve items titled ``senior mobile phones'' for a query ``cellphone for grandpa''.

Traditional web search technology employs the rule-based query rewrite, which transforms the original query to a similar but more standard query, to solve the above hard queries. Figure~\ref{fig:example} illustrates an example of how this query rewriting works in practice.
Those query rewriting rules are normally from a few sources: human compilation, data aggregation and so on. 
However, these rule-based approaches need lots of human efforts that are costly and time-consuming, and they can not cover more subtle cases and long-tail cases. Therefore, there is a significant need for an advanced and powerful system that can solve this problem in a more scalable fashion.

Recently, there is another trend of learning embedding representation to solve this term mismatch problem~\cite{zhang2020towards} and recommendation problems~\cite{mind2019,tree2018}. The basic idea is to mapping queries or users, and item titles to a semantic embedding space, where the queries are close to the relevant items. Therefore, items that do not contain the exact terms but semantically relevant to the query, can be retrieved by the nearest neighbor search in the embedding space. However, in practice, we find this approach suffers from other drawbacks: 1) it is hard to balance the semantic matching ability and too much generalization, which could retrieve irrelevant items. For example, a search query with a very specific intention of a certain model or style of a necklace could retrieve other models or styles of the necklace. 2) It is hard to decide how many items to retrieve from the nearest neighbor search. For some long tail queries with a very specific intention, the number of available and relevant items could be much less than a hyperparameter value, the number of nearest neighbors to retrieve. Thus, the extra retrieved items could cause burdens for the later relevant scoring stages.

In this paper, we will develop a novel approach to the \emph{semantic matching} problem from another perspective of automated and scalable query rewriting. We formulate the query rewriting problem into a cyclic machine translation problem, that first translates query to item titles, and then translates back to queries. To guide the cyclic translation optimally adapt to query rewriting task, we also introduce a novel optimization term to encourage the cyclic translation ``translates back'' to the original query. Our model is flexible enough to leverage most state-of-the-art neural machine translation (NMT) models, \eg, attention-based NMT~\cite{bahdanau2014neural} and transformer-based NMT~\cite{vaswani2017attention}, both of which are based on an encoder-decoder architecture. In practice, we choose the transformer model structure as a skeleton of our query rewriting model, since the transformer has shown superior translation quality and the ability to leverage GPU parallel operations. 

The main contributions of this paper can be summarized as follow

\begin{itemize}
\item In Section~\ref{sec:method}, we present our methods, including models, training and inference algorithms, and system optimizations. Specifically, we develop a novel deep neural network model to query rewriting, which is composed of a cyclic translation formulation of query rewriting in Section~\ref{sec:separate}, a cyclic consistency likelihood in Section~\ref{sec:joint}, an approximated and efficient training algorithm in Section~\ref{sec:training}, an optimal inference algorithm in Section~\ref{sec:inference}, a set of sequence decoding techniques in Section~\ref{sec:decoding}, a few tradeoffs for online serving in Section~\ref{sec:serving} and a few system optimizations in Section~\ref{sec:system}.

\item In Section~\ref{sec:experiment}, we conduct extensive experiments including, ablation studies in Section~\ref{sec:ablation} to clearly illustrate how the models work to generate high quality rewritten queries, offline experiments Section~\ref{sec:offline} to show that cyclic consistency helps improving the performance in, and online A/B experiments in Section~\ref{sec:online} to show that the proposed method is able to improve users' experience significantly.

\item In Section~\ref{sec:discussion}, we conclude our contributions and discuss a few challenging aspects that we have also explored in this query rewriting problem. We will take those directions as our future work, and we also look forward to inspiring more researchers to work on those problems collectively for great practical impacts.
\end{itemize}

\section{Related works}
Our work leverages state-of-the-art neural machine translation models in a novel manner to solve the long existing query rewriting problem, as a counterpart of the embedding based retrieval approach. Thus, we review recent progress in neural machine translation and embedding retrieval. Also, we review classic query rewriting approaches for references.

\subsection{Neural Machine Translation}
Neural Machine Translation (NMT) based on a neural network encoder-decoder architecture recently surpasses its precedent statistical machine translation (SMT) as state-of-the-art for machine translation~\cite{sutskever2014sequence, cho2014learning, wu2016google}. Specifically, the encoder learns a fixed-length embedding as a representation for any token sequences in the source language, which is then used by the decoder to output tokens in the target language.
A few years ago, NMT models are usually based on complex recurrent neural network (RNN), long short-term memory (LSTM)~\cite{hochreiter1997long} and gated recurrent unit (GRU)~\cite{chung2014empirical} to capture the sequential information. Later, attention mechanism~\cite{bahdanau2014neural} between encoder and decoder are introduced to further improve the performance.
More recently, researchers propose to use a pure attention mechanism without any RNN structure, \ie, transformer, to achieve state-of-the-art performance for machine translation~\cite{vaswani2017attention}. The attention mechanism is able to capture global dependencies within the source and the target, and interactions between them. In addition, extra position embedding is added to token embeddings to carry the order information of each token. So far, the transformer has been widely adopted in not only machine translation, but also almost all areas of natural language processing. Several state-of-the-art natural language processing (NLP) models, GPT~\cite{radford2018improving}, GPT2~\cite{radford2019language}, GPT3~\cite{brown2020language} and BERT~\cite{devlin2018bert}, are all based on transformer structure.

Beyond the standard encoder-decoder structure, Tu et al.~\cite{tu2016neural} add a novel reconstructor block to ensure the target sentence contains ``all information'' of the input. Specifically, the reconstructor is trained to translate the target sentence back to the source. The concept of the back-translation is similar to our idea of cyclic translation. This mechanism helps the decoder keep all useful information from the source sentence since otherwise, the reconstructor has an insufficient message to translate back to the source.

\subsection{Embedding Retrieval in Search Engine}
Recently, embedding retrieval technologies have been widely adopted in the modern recommendation and advertising systems~\cite{youtube2016, mind2019, tree2018}, while have not been widely used in search engine yet. We find a few works about retrieval problems in search engine~\cite{palangi2016deep,vu2017search}, while they have not been applied to the industrial production system. DPSR~\cite{zhang2020towards} is one of the first practical explorations in this direction of applying embedding retrieval in the industrial search engine system. Moreover, the well known DSSM~\cite{Huang:2013:LDS:2541176.2505665} and its following work  CDSSM~\cite{Shen:2014:LSR:2567948.2577348} have pioneered the work of using deep neural networks for relevance scoring, which is a very different task from embedding retrieval though.

\begin{figure*}[!ht]
    \centering
    \includegraphics[width=\textwidth]{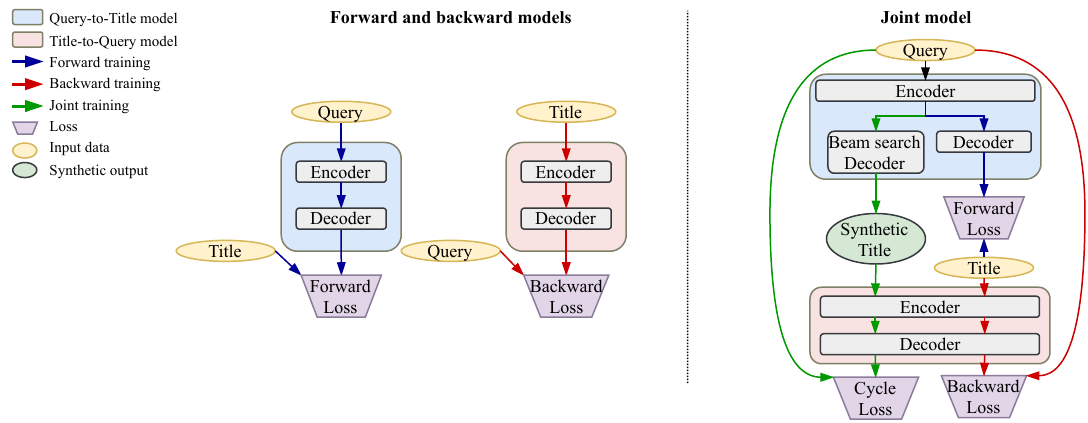}
    \caption{Illustration of the separately trained forward (query-to-title) and backward (title-to-query) models and the jointly trained model.}
    \label{fig:training}
\end{figure*}

\subsection{Classic Approaches to Query Rewriting}
Query expansion, as part of query rewriting, is generally used to formulate a raw query into another one with explicit meaning by adding extra information. In the literature, Bhogal and his colleagues~\cite{bhogal2007review} proposed an ontology-based approach to capture keywords semantics to improve query representation. As shown in another study~\cite{wu2011study}, ontologies may improve the searching performance as an annotated corpus. 
Moreover, thesaurus-based query rewriting is another competitive approach. The company eBay deploys an adaptive approach to generate promising synonym queries, which relies on external resources such as WordNet~\cite{mandal2019query}. These methods heavily rely on the quality of external resources, which makes them not salable to more general circumstances.

The Simrank algorithm~\cite{jeh2002simrank} proposes to construct an object graph to compute similarities. Each edge represents a preference relationship between two objects. The underlying assumption is that similar objects should have common preferences. Thus, similar objects can be identified by the number of shared preferences. As for query rewriting, Antonellis proposed Simrank++~\cite{antonellis2008simrank++} to generate similar queries using user click logs, by assigning weight to each edge according to its total number of clicks. However, this method is not scalable to the current industrial scale of data, which prevents it from widely adopted in modern e-commerce search engine.

More recently, He et al.~\cite{he2016learning} propose a ``learning to rewrite framework'' for the industrial scenario. They take advantage of multiple query rewriters to generate query rewriting candidates, then rank these queries to get the final results. Thus, they do not actually propose a model to generate the rewritten query by itself.

Our approach differs from the above methods in that we train state-of-the-art neural machine translation models, which are capable of leveraging large industrial scale click log data. Apart from the algorithm part, the system of query rewriting is also important.

\section{Method}
\label{sec:method}

In this section, we first introduce an overview of the query rewriting problem and its underlying motivations in Section~\ref{sec:overview}. Then, we develop the proposed method step by step as follows: in Section~\ref{sec:separate} we formulate query rewriting as a cyclic machine translation problem that could already generate reasonable query rewriting results, in Section~\ref{sec:joint} we introduce a novel cyclic consistency loss that can improve the query rewriting performance, in Section~\ref{sec:training} we present an efficient training algorithm that can optimize the originally intractable loss function. Next, we introduce the model inference method as follows: in Section~\ref{sec:inference} we formulate the inference workflow, which, together with an inference trick presented in Section~\ref{sec:decoding}, can generate more diverse query rewriting results.
Finally, we discuss some practical issues that we find particularly important to deploy the proposed method into an industrial system: in Section~\ref{sec:serving} we talk about some tradeoffs that one has to make between inference speed and accuracy to fully deploy the model, and in Section~\ref{sec:system} we introduce another system optimization technique for downstream inverted index retrieval fed with query rewriting results.

\subsection{Overview}
\label{sec:overview}

Query rewriting, which aims at rewriting a given query to another query that can retrieve more relevant items, is a critical task in modern e-commerce search engines. 
The major reason is that item descriptions and user queries are created by different sets of people, who may use different vocabularies and distinct language styles. 
Consequently, even when the queries can perfectly match users’ information needs, the e-commerce search engines may be still unable to locate relevant items.
For example, users who are unfamiliar with Apple's new products, might just type ``newest iPhone'' in an e-commerce website's search input box. However, those relevant items are actually indexed with terms such as ``iPhone 12'' or ``iPhone 12 Pro'' at this moment. 
Ideally, we would like to learn a model which can automatically do the following query rewriting.

\vspace{0.1in}

{
\centering
\begin{tabular}{cccc}
    $\quad\quad\quad$&\textrm{newest iPhone}  &  & \textrm{iPhone 12} \\
    &$\uparrow$ & & $\uparrow$ \\
    &\textrm{model input} & & \textrm{model output}
\end{tabular}
}
\vspace{0.1in}


The straightforward way to learn a query-to-query translation model from a given data set of enough query rewriting logs. As one of the largest e-commerce platforms, we still can not collect a sufficient number of high-quality query rewriting logs, since the raw query rewriting logs are not of guaranteed quality. Thus, human labeling or rule-based methods are necessary to extract a small number of high-quality data, which becomes too expensive and impractical to train deep learning models. Thus, to overcome this obstacle, we have to resort to some other approaches to query rewriting.

One can easily find out that the availability of a tremendous amount of click log data could be a rescue to the problem, though the click logs are actually query-to-title data. How to utilize this type of data to learn the query-to-query translation model would be an interesting and potentially promising direction to explore. Basically, as illustrated in the left part of Figure~\ref{fig:training}, the intuition is that we can train two translation models, the forward (query-to-title) model and the backward (title-to-query) model, to accomplish this task. However, two separately trained translation models are not going to be optimal in terms of generating the best rewritten queries. 
In Section~\ref{sec:joint}, we will talk about how to jointly train the two translation models to get the optimal rewritten queries.

\subsection{Query Rewriting As a Cyclic Translation}
\label{sec:separate}
Given a click log data $\D = \{\x^{(n)}, \y^{(n)}\}_{n=1}^{N}$, where $\x$ denotes query, $\y$ denotes item title, and $N$ denotes the number of training samples, the standard training objective in most translation models is to maximize the log likelihood of the training data
\begin{align}
L_f(\thetaf) &= \sum_{n=1}^{N}\log P(\y^{(n)} |\x^{(n)};\thetaf), \label{eq:loss_f}\\
L_b(\thetab) &= \sum_{n=1}^{N}\log P(\x^{(n)} |\y^{(n)};\thetab), \label{eq:loss_b}
\end{align}
where $P(\y |\x;\thetaf)$ and $P(\x |\y;\thetab)$ are query-to-title (forward) and title-to-query (backward) neural translation models, parameterized by $\thetaf$ and $\thetab$, respectively. Note that the subscripts $f$ and $b$ are shorthands for \textrm{forward} and \textrm{backward}, respectively.
Two objective functions $L_f$ and $L_b$ are independent on each other. Thus, the model can be trained separately without loss of accuracy. 

In practice, we find that the query-to-title model requires more memorization capability in order to generate good enough item titles, potentially due to that target item titles are normally much longer than source queries and potentially item title space is much larger than query space. Thus, the translation model has to ``memorize'' the item titles to generate corresponding titles for a given query. On the other hand, we find that the title-to-query model is more like a text summarization model. Thus, it does not require a large model size to memorize. 

Our query rewriting model is general enough to leverage most neural machine translation (NMT) models for both forward and backward directions. We have experimented with attention-based model~\cite{bahdanau2014neural} and transformer-based model~\cite{vaswani2017attention}, both of which work well in our scenario but the latter shows slightly better performance (see Experiments in Section~\ref{sec:translation_models}). Thus, we choose a $4$-layers transformer for the query-to-title model, and $1$-layer transformer for the title-to-query model. The detailed model setup can be found in Section~\ref{sec:setup}.

After the models are trained, we can simply run the two models sequentially to generate a rewritten query, via intermediate synthetic item titles, as illustrated in Figure~\ref{fig:inference}. We will present a more disciplined approach in Section~\ref{sec:inference}.

\subsection{Cyclic Consistency}
\label{sec:joint}
The above two separately trained models work reasonable well in practice (see Table~\ref{tab:good_examples_separate}, Table~\ref{tab:comppare with baseline}). However, as one can imagine, it is sub-optimal in terms of generating the best query rewriting, since the learning algorithm does not specifically take the task, query rewriting, into account. 

Our idea is to leverage a \emph{cycle consistency} in learning the two models. For our mission, we have two translation models, query-to-title and title-to-query. To get a better query rewriting model, the intuition is to encourage the two translation models can collaboratively ``translate back'' to the original query.
Thus, model parameters should be learned to maximize the likelihood of ``translating back'' the original query.

Formally, we introduce a cycle consistent likelihood $L_c(\thetaf, \thetab)$ to encourage two models collaboratively ``translate back'' the original query as follows
\begin{align}
L_c(\thetaf, \thetab) &= \sum_{n=1}^{N} \log P(\x^{(n)} | \x^{(n)}; \thetaf, \thetab) \nonumber \\
&= \sum_{n=1}^{N} \sum_{\y \in \Y} \log  P(\y|\x^{(n)};\thetaf) P(\x^{(n)}|\y;\thetab),
\label{eq:loss_c}
\end{align}
where for each training query $\x^{(n)}$, it computes the ``translating back'' probability $ P(\x^{(n)} | \x^{(n)})$ by marginalize over all possible item titles $\y$. 

The final likelihood function is a linear combination of forward, backward and cyclic consistency likelihoods in Equations~(\ref{eq:loss_f}), (\ref{eq:loss_b}) and (\ref{eq:loss_c}) as follows
\begin{equation*}
L(\thetaf, \thetab) = L_f(\thetaf) + L_b(\thetab) + \lambda L_c(\thetaf, \thetab),
\end{equation*}
where the hyper-parameter $\lambda$ controls the tradeoff between the bi-directional translation likelihood and the cycle consistent likelihood.

Thus, the optimal model parameters, $\btheta^*_f$ and $\btheta^*_b$, are learned by
\begin{align*}
    \btheta^*_f=\arg\max  & \left\{\sum_{n=1}^{N}\log P(\y^{(n)} |\x^{(n)};\thetaf) + \right. \nonumber \\
                          & \left.\lambda \sum_{n=1}^N\log P(\x^{(n)}|\x^{(n)};\thetaf,\thetab)  \right\},
\end{align*}
\begin{align*}
    \btheta^*_b=\arg\max & \left\{\sum_{n=1}^{N}\log P(\x^{(n)} |\y^{(n)};\thetab) + \right.\nonumber \\
                         & \left.\lambda \sum_{n=1}^N\log P(\x^{(n)}|\x^{(n)};\thetaf,\thetab) \right\}.
\end{align*}
We can see that the query-to-title and the title-to-query models are connected via the cyclic consistency and they can hopefully benefit each other in joint training.

\subsection{Training}
\label{sec:training}

The partial derivative of $L(\thetaf, \thetab)$ with respect to the forward model parameter $\thetaf$ can be written as follows. 
\begin{align}
\frac{\partial L(\thetaf, \thetab)}{\partial \thetaf} 
&= \sum_{n=1}^{N} \frac{\partial{\log P(\y^{(n)} |\x^{(n)};\thetaf)}}{\partial{\thetaf}} \nonumber \\
&+ \lambda \sum_{n=1}^N 
\frac{\partial{\log \sum_{\y_i \in \Y} P(\y_i|\x^{(n)};\thetaf) P(\x^{(n)}|\y_i;\thetab) }}{\partial{\thetaf}}. \label{eq:grad}
\end{align}
We skip the partial derivative for the backward model for brevity, which can be derived similarly. 

However in practice, it is prohibitively expensive to compute the sums in Equation~(\ref{eq:grad}) due to the exponential search space of $\Y$. Alternatively, we propose to use a subset of the full space $\tilde{\Y} =  \{{\y_1,\y_2,\cdots,\y_k}\} \subset \Y$ to approximate the second term in Equation~(\ref{eq:grad}) as follows
\begin{align}
&\sum_{n=1}^{N} \frac{\partial \log  \sum_{\y_i \in \tilde{\Y}} P(\y_i|\x^{(n)};\thetaf) P(\x^{(n)}|\y_i;\thetab)}{\partial \thetaf} \nonumber \\
=& \sum_{n=1}^{N}  \frac{ \sum_{\y_i \in \tilde{\Y}} P(\x^{(n)}|\y_i;\thetab) 
\cdot \frac{\partial P(\y_i|\x^{(n)};\thetaf)}{\partial \thetaf}}
{ \sum_{\y_i \in \tilde{\Y}} P(\y_i|\x^{(n)};\thetaf) P(\x^{(n)}|\y_i;\thetab)}.
\label{eq:loss_c_approx}
\end{align}
In practice, we use the top-$k$ set of item titles $\y_i$ translated by the forward model for the given query $\x^{(n)}$. As $k \ll |\Y|$, it is feasible to calculate Equation~(\ref{eq:loss_c_approx}) efficiently by enumerating all $\y_i$ in $\tilde{\Y}$. 

The subset $\tilde{\Y}$ can be obtained by any sequence decoding methods (see Section~\ref{sec:decoding}) with the forward model $P(\y|\x;\thetaf)$. In practice, we find this step is much more time consuming than other steps since we have to run the decoder network for several steps according to the item title length. Also, consider that the cyclic consistency only makes sense when the two models are well trained. Thus, we only perform the cyclic consistency term in Equation~(\ref{eq:loss_c_approx}) after a certain number of warmup steps. We use Adam optimizer~\cite{kingma2014adam} for the training steps. The detailed algorithm is shown in Algorithm~\ref{alg:joint}. We also illustrate the training process visually in Figure~\ref{fig:training}.

{\begin{algorithm}
\begin{algorithmic}[1]
\STATE \textbf{input}: Dataset $D$=$\{\x^{(n)}, \y^{(n)}\}_{n=1}^{N}$, batch size $B$, max steps $T$, beam width $k$, warmup training steps $G$.
\STATE Random initialize model parameters as $\thetaf^{(0)}$ and $\thetab^{(0)}$.
\FOR{$t = 1 \ldots T$}
\STATE Sample a batch of $b$ examples $\mathcal{B}=\{\x^{(b)}, \y^{(b)}\}_{b=1}^{B} \subseteq \mathcal{D}$.
\IF{$t \le G$}
\STATE Compute two model gradients $\frac{\partial{L_f}}{\partial{\thetaf^{(t-1)}}}$ and $\frac{\partial{L_b}}{\partial{\thetab^{(t-1)}}}$.
\STATE Update model parameters $\thetaf^{(t)}$ and $\thetab^{(t)}$ by Adam optimization step.
\ELSE
\STATE Perform top-$n$ sampling with forward model $P(\y| \x; \thetaf^{(t)})$ to generate $k$ synthetic titles for each query.
\STATE Collect synthetic data set $\left\{\{\x^{(b)}, \y^{(b)}_i\}_{i=1}^{k}\right\}_{b=1}^B$.
\STATE Compute two model gradients $\frac{\partial{L}}{\partial{\thetaf^{(t-1)}}}$ and $\frac{\partial{L}}{\partial{\thetab^{(t-1)}}}$.
\STATE Update model parameters $\thetaf^{(t)}$ and $\thetab^{(t)}$ by Adam optimization step.
\ENDIF
\ENDFOR
\end{algorithmic}
\caption{Cyclic Consistent Training Algorithm}
\label{alg:joint}
\end{algorithm}}

\subsection{Inference}
\label{sec:inference}

After the forward and backward models are trained, we obtain the optimal model parameters $\thetaf^*$ and $\thetab^*$. Then the inference problem can be formulated as follows. Given a query $\x$, we would like to find another query $\x^* \neq \x$ that maximizes the following probability
\begin{align*}
x^* &= \arg\max_{x'} P(\x'|\x; \thetaf^*, \thetab^*) \nonumber \\
&= \arg\max_{x'} \sum_{\y\in \Y} P(\y | \x; \thetaf^*) P(\x' | \y; \thetab^*).
\end{align*}
However, it is infeasible to enumerate all possible item title $\Y$ for the sum. Therefore, again, we have to resort to a subset $\tilde{\Y} \subset \Y$ to approximate the inference. 

\begin{figure*}
    \centering
    \includegraphics[width=\textwidth]{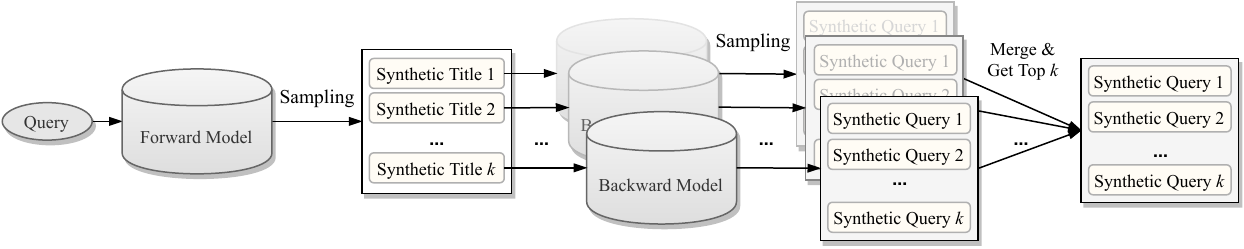}
    \caption{Inference process of generating top synonymous queries for one user input query.}
    \label{fig:inference}
\end{figure*}

The full workflow of inference is shown in Figure~\ref{fig:inference}. Specifically, we perform top-$n$ sampling decoding method (detail is described in \ref{sec:decoding}), where $n$ means the number of candidate token ids, for a given query $\x$ with the forward model to generate $k$ synthetic item titles in $\tilde{\Y} = \{\y_1, \ldots, \y_k\}$, according to translation probabilities $P(\y_i|\x; \thetaf^*)$. Then, we perform top-$n$ sampling  for each item title $\y_i$ with the backward model to generate $k$ synthetic queries each, denoted as $\x_{ij}$ where $1\le j \le k$, according to translation probabilities $P(\x_{ij} | \y_i; \thetab^*)$. Finally, we get the most likely $k$ synthetic queries from the candidate set $\{\x_{ij}| i, j\}$ of size $k^2$, according to the probability
\begin{align*}
P(\x_{ij} | \x) = \sum_{1\le t \le k} P(\y_t|\x; \thetaf^*) P(\x_{ij} | \y_t; \thetab^*).
\end{align*}
Note that in practice, all the above computations are performed in log probability space, and we have to carefully apply a few tricks to avoid numerical issues~\cite{nielsen2016guaranteed}. 



\subsection{Diverse Sequence Decoding}
\label{sec:decoding}

As we mentioned before in Equation~(\ref{eq:loss_c_approx}), the top-$k$ item titles in set $\tilde{\Y}$ are obtained by any sequence decoding algorithm. 
Generally speaking, the optimal sequence, \ie, the most likely sequence, can only be obtained by exhaustive search, which is infeasible for most cases. In practice, typically, greedy search and beam search are the two most widely used approaches. In detail, the greedy search selects the most likely token at each step of the decoding process. Thus, it is not guaranteed to find the optimal sequence, as the global most likely sequence might not locally take the most likely token at each step. Beam search is an improved algorithm based on the greedy search. Instead of picking the most likely token at each step, the beam search maintains a number (\ie, beam size) of most likely sequences during the decoding process. More details about these two sequence decoding methods can be found in~\cite{bisiani1987encyclopedia}.

In practice, we found the above two widely used methods are not well applicable to our problem. Greedy search outputs only one sequence, which does not fit into our algorithm. Beam search, however, outputs very similar sequences that lack diversity. For example, we found some synthetic item titles only differ in a blank space, or a single token. Thus, those almost identical item titles lead to very similar query rewriting results, which is not what we want in practice.

\begin{figure}
    \centering
    \includegraphics[width=0.45\textwidth]{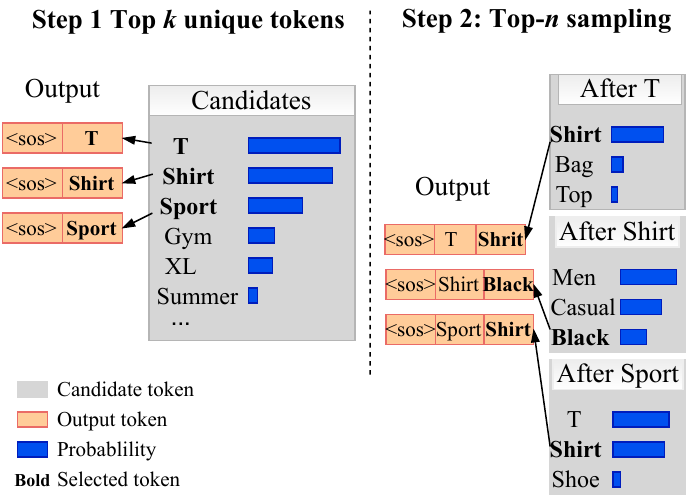}
    \caption{Illustration of the top-$n$ decoding method. At the first step, the most likely $k=3$ unique tokens (``T'', ``Shirt'', ``Sport'') are selected. In the following steps, for each candidate sequence, the top $n=3$ most likely tokens are selected as candidate set, then we sample one token from them according to their probability.}
    \label{fig:decoding}
\end{figure}

Therefore, we develop a novel sequence decoding technique, namely top-$n$ sampling decoder, to generate more diverse sequences. As shown in Figure~\ref{fig:decoding}, we start by maintaining $k$ candidate sequences. At the first step, we pick the most likely $k$ tokens to ensure all the candidate sequences begin differently. This is a key step to increase the result's diversity. At the following steps, we perform top-$n$ sampling to obtain the tokens for each candidate sequence. Specifically, we do sampling among the top $n$ ($n$ = 40 in our experiments) most likely tokens, according to their conditional generative probability (\ie, the softmax layer output in the neural network). Thus, this top-$n$ sampling could well balance the overall likelihood and diversity in the sequence output.

\subsection{Online Serving}
\label{sec:serving}

Even though the proposed model is capable to generate reasonably good query rewriting, in practice we found it is still very challenging to deploy the model inference online. The typical total model inference latency for the query-to-title and title-to-query translations is more than 100 milliseconds, even in a modern GPU machine. Thus, it is significantly beyond the typical industrial backend system latency requirement, which is normally $50$ milliseconds. 
 
At the first step, we run the above proposed model offline to generate query rewriting for top $8$ million popular queries, which are then fed into an online key-value store for fast online retrieval (less than 5 milliseconds). Those queries cover more than $80\%$ 
of our search engine traffic. To cover all queries, especially for long tail queries that need query rewriting more than top queries, we have to speed up the proposed approach by trading off some accuracy potentially. 

In practice, we make two modifications to speed up the model inference. First, we notice that the most time-consuming steps are the two sequence decodings, from query to title and from title to query. Thus, we simplify it by performing a query to query translation directly. The training data is prepared by collecting queries that share more than a certain number of clicks to the same items. We take these query pairs as synonymous queries. Then we train a single translation model on these synonymous query pairs, in order to reduce the sequence decoding time. Second, we notice that the transformer layer is more powerful but more time consuming than the traditional Recurrent Neural Network (RNN) layer. Especially, the transformer decoder is the bottleneck for the whole model inference, since the multi-head self attention~\cite{vaswani2017attention} needs to be performed for all target tokens at each decoding step. Its counterpart RNN decoder, however, is much cheaper, since it only takes constant computing time at each decoding step.
For the encoder, we still keep the transformer encoder for better accuracy (see Figure~\ref{fig:hybrid_rnn}). Eventually, we are able to reduce the model inference time to about 30 milliseconds on a typical industrial CPU machine with 32 physical cores, which can be deployed online to handle all query rewritings.

\subsection{System Optimization by Merging Syntax Trees}
\label{sec:system}

The query rewriting generates a few more queries in addition to the original query, which incurs extra burdens and challenges to the retrieval system, especially to the inverted index retrieval where the majority of computation happens during retrieval. In this section, we will explain how we optimize the inverted index retrieval to make the proposed method feasible in practice.

\begin{figure}
    \centering
    \includegraphics[width=0.4\textwidth]{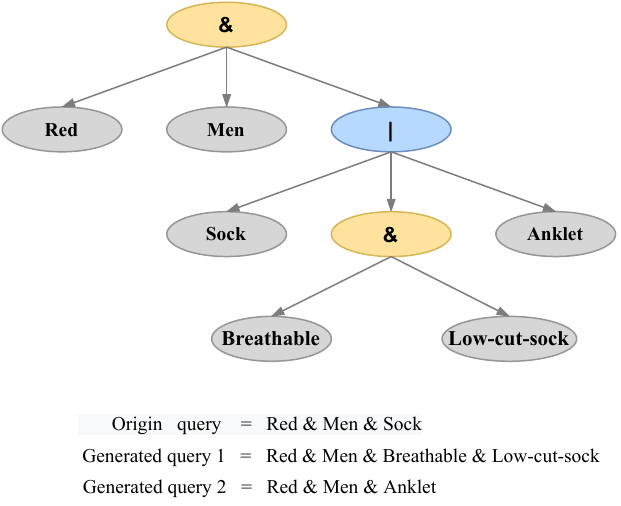}
    \caption{Merged syntax tree for two generated queries and one original query. The node ``\&'' stands for logical ``and'' operation, and the node ``$\mid$'' stands for ``or'' operation for inverted index retrievals.}
    \label{fig:system_sq}
\end{figure}

Given a user input query, our search engine first constructs a syntax tree by text tokenization and syntactic analysis, which is then used to extract document lists from the inverted index. The most straightforward way to implement the proposed query rewriting system is to construct as many syntax trees as the number of rewritten queries. This looks reasonable in theory. However, in practice, we find this straightforward approach is unfortunately inefficient, in terms of much more CPU usage and much longer system latency than the previous one-query-retrieval. To overcome this technical challenge in the system, we optimize the syntax tree construction to still keep only one tree by merging all the rewritten queries and the original query. 
The merged tree is much smaller than the sum of all syntax trees built separately for each query, since there are many common tokens between rewritten queries and the original query. As a result, we are able to keep the merged tree for multiple queries slightly larger than the previous tree for only the original query. This significantly reduces the retrieval system computation cost, in order to make the proposed model practical.  Figure~\ref{fig:system_sq} shows an example of how we can merge two generated queries and an original query into one syntax tree. Typically, we use a \emph{or} operation (denoted as ``$|$'' node) to merge all possible tokens if they diverge at the position.

\section{Experiments}
\label{sec:experiment}

\subsection{Setup}
\label{sec:setup}
We use $60$ days user click logs as our training data set, and only keep those samples with more than one click. Since only one click over a period of two months is very likely to be an accidental and unintentional click, which could pollute the quality of the dataset. The statistics of the dataset is shown in Table~\ref{tab:data}. 

\begin{table}[h]
    \caption{Statistics of data set}
    \label{tab:data}
    \centering
    \begin{tabular}{c|c|c}
    \hline 
    & \textbf{Query} & \textbf{Item Title}\\
    \hline
    \# Query Item Pairs     &  \multicolumn{2}{c}{300 million} \\
    \hline
    \# Search Sessions & \multicolumn{2}{c}{5.6 billion} \\
    \hline
    Vocab Size & \multicolumn{2}{c}{9744} \\
    \hline
    \# Average Words     &  6.12 &  49.96\\
    \hline
    \end{tabular}
    
\end{table}

We tried a few model structures to balance the training speed and model performance. The final model is composed of two different transformers, whose detail structures are shown in Table~\ref{tab:hyer-param}.
Moreover, we set hyper-parameter $\lambda=0.1$, beam width $k=3$, and applied Adam optimizer~\cite{kingma2014adam} with learning rate=$0.05$, $\beta_1=0.9$, $\beta_2=0.999$ and $\epsilon=10^{-8}$. Noam scheduler~\cite{vaswani2017attention} that varies learning rate according to training steps is adopted in the optimization.

\begin{table}[h]
    \caption{Model Hyperparameters}
    \label{tab:hyer-param}
    \centering
    \begin{tabular}{c|c|c}
    \hline 
    & \textbf{Query-to-title} & \textbf{Title-to-query}\\
    \hline
    \# Transformer Layer & 4 & 1\\
    \hline
    \# Head & \multicolumn{2}{c}{8} \\
    \hline
    Hidden Units of Feed-forward & \multicolumn{2}{c}{1024} \\
    \hline
    Embedding Dimensionality     &
    \multicolumn{2}{c}{512}\\
    \hline
    Dropout Rate & \multicolumn{2}{c}{0.1}\\
    \hline
    \end{tabular}
    
\end{table}

\begin{CJK*}{UTF8}{gbsn}
\begin{table*}[t]
    \caption{Good cases from separately trained models}
    \label{tab:good_examples_separate}
    \centering
    \begin{tabular}{|c|c|c|}
    \hline
        \textbf{Original Query} &  \textbf{Synthetic Item Title} & \textbf{Rewritten Query}\\
    \hline
        给爷爷的手机 &Second hand Apple 8plus mobile phone apple iphone8plus  &  苹果8plus \\
        (cellphone for grandpa) & apple 8p golden 64g full Netcom & (iphone 8plus) \\
    \hline
        老人奶粉 &
        Brand 'Yili' milk powder golden collar crown series infant formula  &  金领冠3段 \\
        (milk powder for seniors) & milk powder 3 segments 900g*3 cans & (kido level-3)\\
    \hline
        自营猪年纪念币 &
        Chinese Gold Coin 2020 Year of the Rat Commemorative Coin 2010 & 鼠年纪念币 \\
        (commemorative coins & Year of the Rat Commemorative Coin 2010 & (commemorative coins \\
        for Year of the Boar)&  & for Year of the Rat)\\
    \hline
        男士去皱 & Nivea Men's Eau De Toilette 50ml+Men's Perfume 50ml+ & 男士香水 \\
        (wrinkle removal for men)& Men's Eau De Toilette 50ml+Men's Eau De Toilette 50ml & (men's perfume)\\
    \hline
    \end{tabular}
\end{table*}
\end{CJK*}

\begin{CJK*}{UTF8}{gbsn}
\begin{table*}[t]
    \caption{Good cases from jointly trained model}
    \label{tab:good_examples_joint}
    \begin{tabular}{|c|c|c|}
    \hline
        \textbf{Original Query} &  \textbf{Synthetic Item Title} & \textbf{Rewritten Query}\\
    \hline
        给爷爷的手机 &
        Little Pepper Mobile Phone Full Netcom 4g Dual SIM Dual Standby Mobile Phone for the  &  老人手机 \\ (cellphone for grandpa)& Elderly Mobile Unicom 2g Dual SIM Mobile Phone for the Elderly, Student Standby Black
        &(senior phone) \\
    \hline
        老人奶粉 &
        Imported from New Zealand Anjia anchor whole  &  奶粉成人 \\(milk powder for seniors)
        & milk powder adult skimmed milk powder 1kg bag & (adult milk powder) \\
     \hline
        自营猪年纪念币 &
        2019 Year of the Pig Zodiac Commemorative Coin Second Round & 猪年纪念币 \\
        (commemorative coins & of Zodiac Circulation Commemorative Coin 10 Yuan Face Value & (commemorative coins\\
         for Year of the Boar)&& of for Year the Boar)\\
    \hline
        男士去皱 & L'Oreal loreal men's sharp anti-wrinkle firming and diminishing fine & 男士护肤品套装 \\
        (wrinkle removal for men)& lines moisturizing facial skin care cosmetics set authentic five-piece set & (men's skin care set)\\
    \hline
    \end{tabular}
\end{table*}
\end{CJK*}

\subsection{Ablation Study}
\label{sec:ablation}

\subsubsection{Examples}
In Table~\ref{tab:good_examples_separate} and~\ref{tab:good_examples_joint}, we show some examples of the query rewriting results. As we can see from the table, the proposed models are able to work surprisingly well for some hard queries, such as ``cellphone for grandpa'' to ``senior cellphone'', ``milk powder for seniors'' to ``milk powder for adults''. Also, we can observe that the jointly trained model performs better than the separately trained model (without the cyclic likelihood). These examples have already shown the difference.

\begin{CJK*}{UTF8}{gbsn}
\begin{figure*}[!ht]
    \centering
    \includegraphics[width=1.03\textwidth]{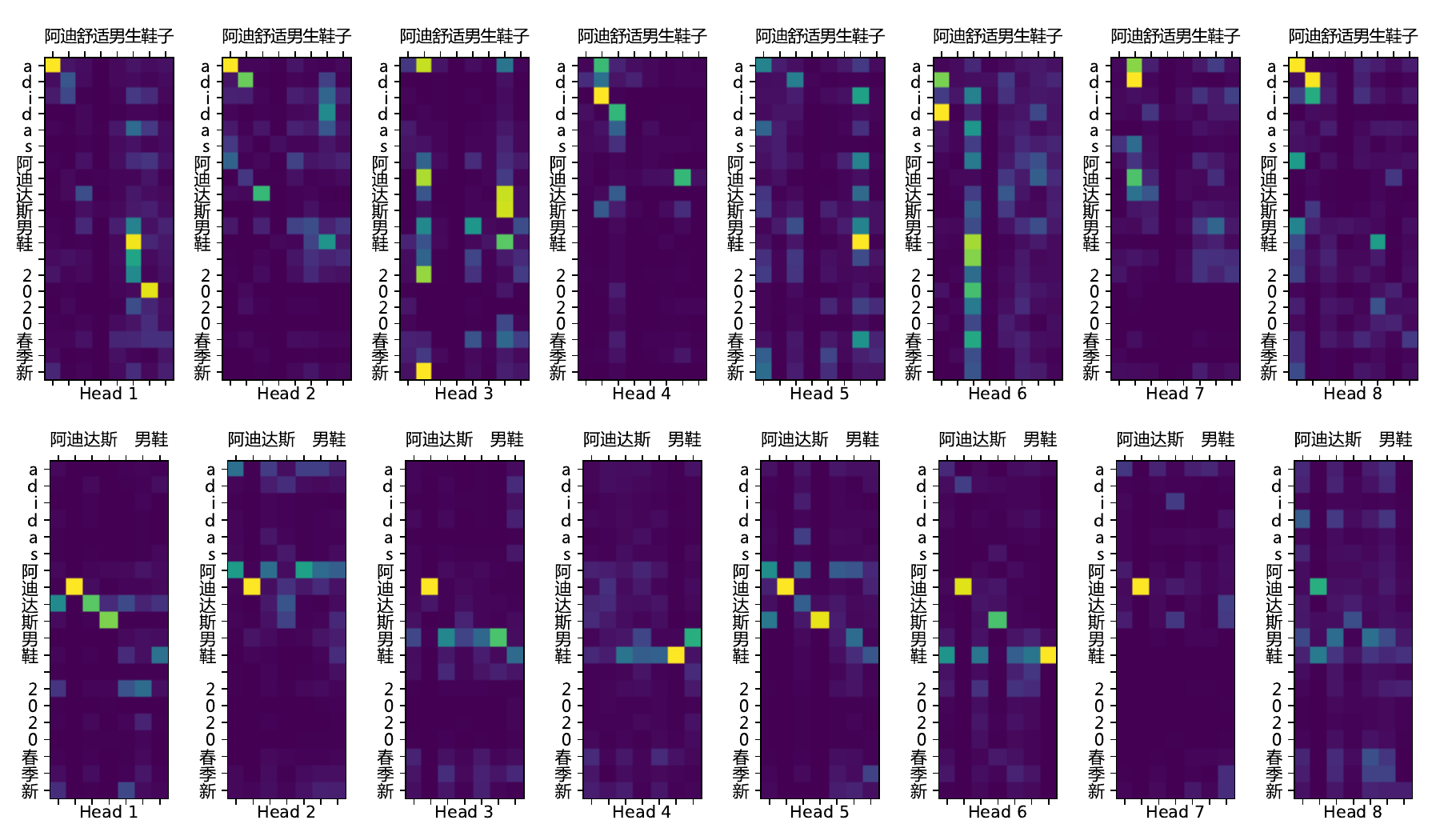}
    \caption{Heat map of attention weights between query and synthetic titles (above), and between synthetic titles and rewritten query (below). The x axis corresponds to query and the y axis corresponds to synthetic title. The brightness of block represents the attention weights. Chinese (English) translations are 阿迪达斯(Adidas), 舒适(comfort), 男生(men's), 鞋子(shoe), 男鞋(men's shoe)，春季(spring), 新(new).}
    \label{fig:attention}
\end{figure*}
\end{CJK*}

\subsubsection{Attention Visualization}
To have a better idea of how the attention-based translation works for our models, we illustrate in Figure~\ref{fig:attention} a heat map of the attention weights of the two translation steps, from query to synthetic titles, and from synthetic titles to rewritten query. The example rewrites ``Ah Di comfortable men's shoe'', that is composed of a shorthand ``Ah Di'' for Adidas in Chinese and a vague descriptive word ``comfortable'', to ``Adidas men's shoe'' which is more standard for the retrieval task. From the attention weights, we can see that the query-to-title model is able to let ``Ah Di'', a shorthand for Adidas in Chinese attend the English brand name ``adidas'', to let ``male student's shoe'' attend ``men's shoe'' and to skip the vague descriptive word ``comfortable''. Then, the title-to-query model is able to leverage the other full Chinese brand name for Adidas in title to be translated into a rewritten query. This is just an example to illustrate how the proposed models work in practice to generate more clean and standard queries for a vague user input query.

\begin{figure}[!t]
    \centering
    \subfigure[Perplexity]{
    \begin{minipage}[]{0.5\textwidth}
        \centering
        \includegraphics[width=\textwidth]{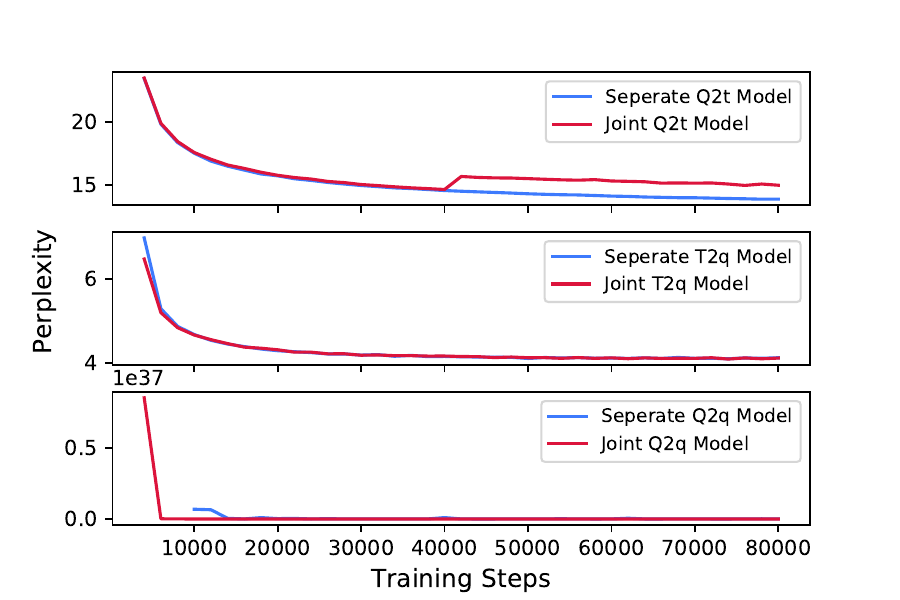}
        \label{fig:perplexity}
    \end{minipage}
    }
    \subfigure[Log probability]{
    \begin{minipage}[]{0.5\textwidth}
        \centering
        \includegraphics[width=\textwidth]{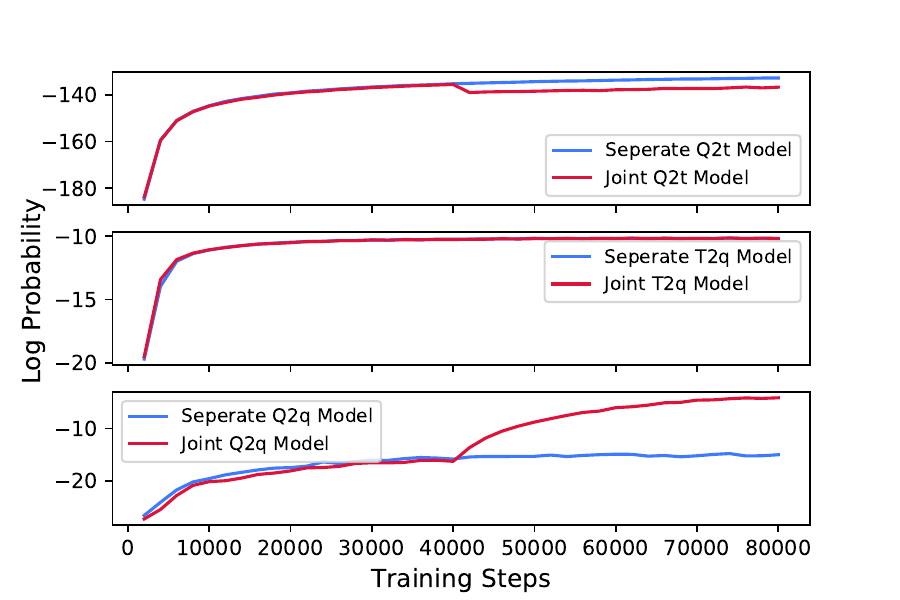}
        \label{fig:log_prob}
    \end{minipage}
    }
    \subfigure[Accuracy]{
    \begin{minipage}[]{0.5\textwidth}
        \centering
        \includegraphics[width=\textwidth]{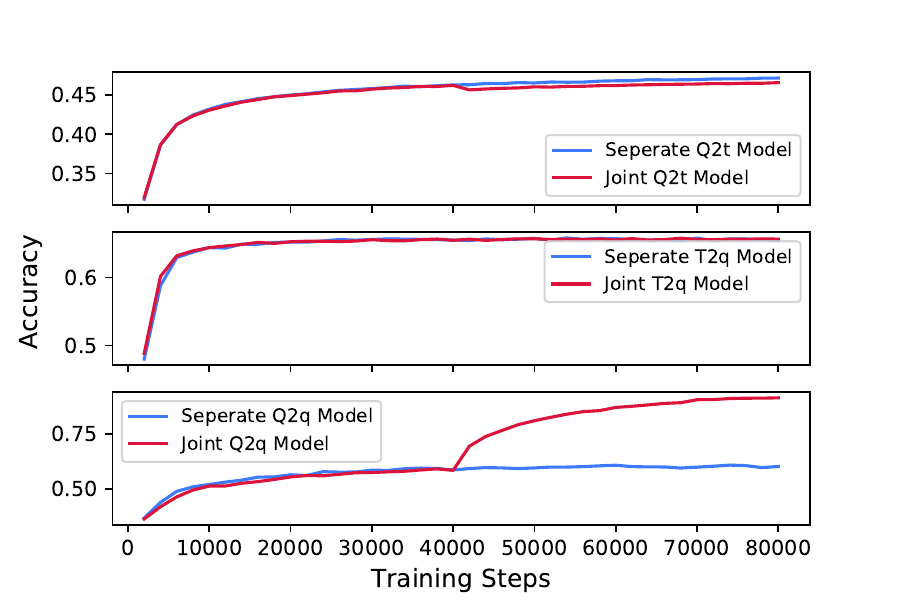}
        \label{fig:accuracy}
    \end{minipage}
    }
    \caption{Comparison of training convergence curve between separately trained models and jointly trained models.}
    \label{fig:main_metrics}
\end{figure}

\subsubsection{Training Convergence}
\label{sec:convergence}
We compared the model from jointly training and separately training. The below metrics are used for evaluations.

\begin{itemize}
    \item Perplexity is a widely used metric in natural language processing, which measures how well a probability model predicts the training label. The value of perplexity can be computed by an exponential of cross entropy loss. Thus, the smaller perplexity is, the better the model performance is.
    \item Log probability stands for the log of the probability of ``translate back'' the original query, by marginalizing over a fixed number of intermediate synthetic titles, that are sampled using top-$n$ sampling method. The higher log probability is, the better the model performance is.
    \item Accuracy is a similar metric to log probability. Instead of computing the probability of ``translate back'' the original query, we compute the accuracy of predicting the same token as the original query at each position.
\end{itemize}

Figure~\ref{fig:main_metrics} shows the comparison between separately trained query-to-title (q2t) and title-to-query (t2q) models, and jointed trained ones. We can see that there is a significant jump on all metrics after $40,000$ warm-up steps, when the joint training starts adding the cyclic likelihood. 


Also, we notice that the joint training does not affect the quality of title-to-query translation, since all the metrics keep the same. The query-to-title translation is slightly affected, presumably as a tradeoff for better query-to-query accuracy. The comparison demonstrates that the joint training with the cyclic likelihood could significantly boost the performance of query-to-query translation, as well as the query rewriting quality.

\subsection{Offline Experiments}
\label{sec:offline}
In this section, we talk about a few offline experiments that support our choice of translation models, human evaluation of query rewriting relevancy, and comparison with baseline methods.

\subsubsection{Choices of Translation Models}
\label{sec:translation_models}

\begin{figure}[tb]
    \centering
    \includegraphics[width=0.5\textwidth]{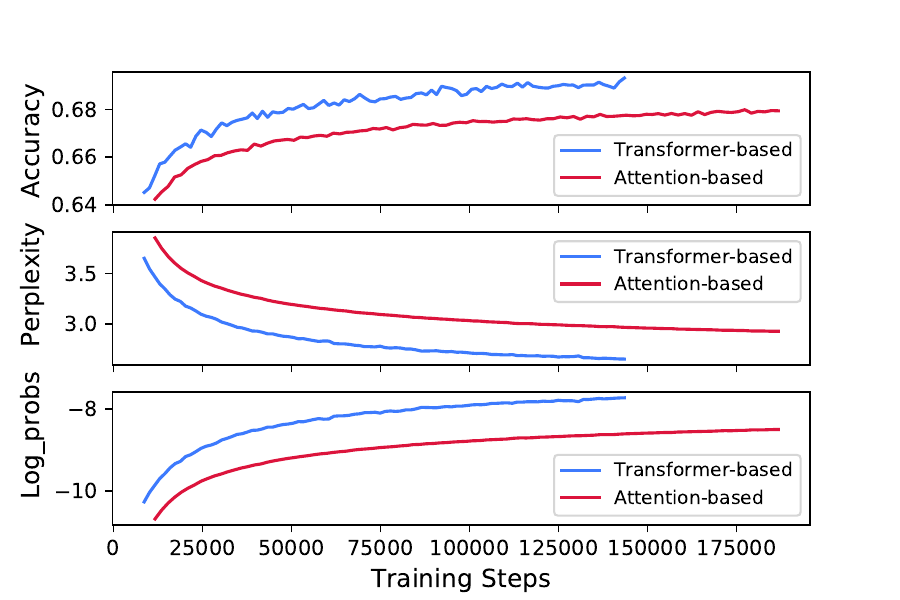}
    \caption{Comparison between transformer-based and attention-based methods in our scenario.}
    \label{fig:trans_atten_compare}
\end{figure}

Our proposed model is flexible in using different machine translation algorithm. We compare two most widely used model structures, transformer-based~\cite{vaswani2017attention}, and attention-based~\cite{bahdanau2014neural} ones, in our scenario. Figure~\ref{fig:trans_atten_compare} shows the comparison results. It is clear that the transformer-based model provides significantly better results than the attention-based model on all three metrics. However, we also notice that the transformer model requires more computations.
Thus, we use a transformer-based machine translation model exclusively for all our experiments, if not specified otherwise.

\begin{figure}[t]
    \centering
    \includegraphics[width=0.5\textwidth]{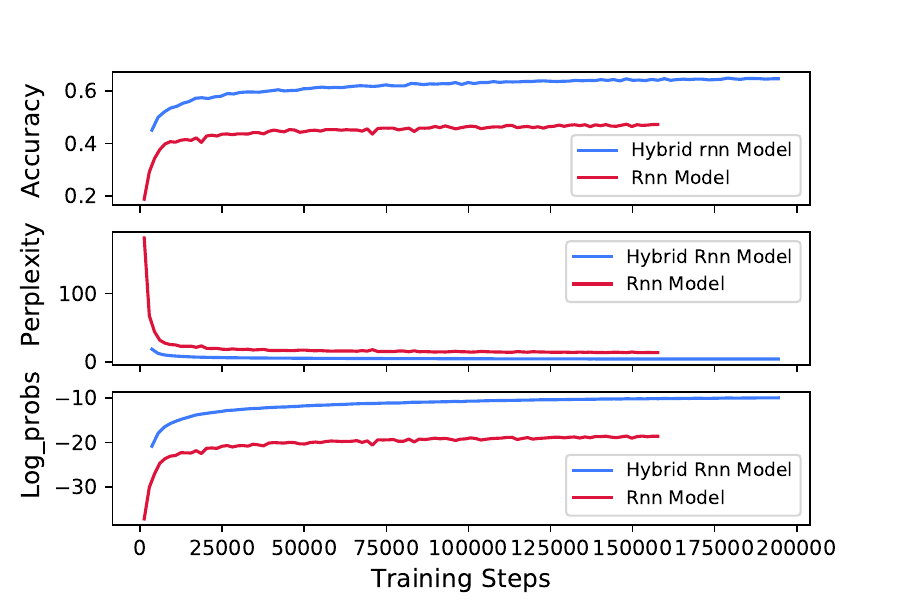}
    \caption{Comparison between RNN and Hybrid RNN on direct query-to-query training.}
    \label{fig:hybrid_rnn}
\end{figure}

As discussed in Section~\ref{sec:serving}, we have to simplify the model to deploy it online. We test the latency performance on different types of encoders and decoders on CPU with the same parameters: beam width as 3, the layer of encoder and decoder as 1, vocabulary size as 3000. and the maximum decode step as 15. As shown in Table~\ref{tab:latency}, transformer encoder and RNN decoder get the best results. Therefore, we consider two types of simplification: a pure RNN based model that uses RNN for both encoder and decoder, and a hybrid RNN model that uses RNN for decoder only. Figure~\ref{fig:hybrid_rnn} shows a comparison between these two types of models. The hybrid RNN model shows significantly better results than the pure RNN model, which indicates that the transformer encoder is still necessary for a balance between query rewriting quality and online serving latency.

\begin{table}[t]
    \caption{Latency in milliseconds of different translation models}
    \label{tab:latency}
    \centering
    \begin{tabular}{c|ccc}
    \hline 
    & \textbf{RNN} & \textbf{GRU} & \textbf{Transformer}\\
    \hline
    Encoder & $6$ & $9$ & $3.5$ \\
    Decoder & $30$ & $35$ & $67.5$ \\
    \hline
    \end{tabular}
\end{table}

\subsubsection{Human Evaluation of Relevancy}
\label{sec:human_labeling}
Since rewritten query relevancy is hard to evaluate in an automatic way, we resort to human labeling for the evaluation. 
We first randomly select $1,000$ queries as our evaluation set which also have rule-based synonyms. Then we generate three rewritten queries for each query in the evaluation set using a separately trained model and jointly trained model respectively.
As Table~\ref{tab:human_eval} shows, human labelers are asked to evaluate twice on two comparisons, joint model versus separate model, and joint versus rule-based method. The result demonstrates that the jointly trained model generates more relevant rewritten queries than the separately trained model, which indicates the effectiveness of our joint training algorithm. We also compare the jointly trained model with the rule-based method for the same data set. It is expected that the rule-based method is more reliable on relevance since it often only replaces a single word in the query by lookup to a human curated dictionary. 
Surprisingly, the jointly trained model still wins in some cases of disambiguation of polysemous words. For example, the word ``cherry'' may be replaced with its synonym by a rule-based dictionary. However, the synonym describes, in fact, the ``cherry'' brand. In contrast, the joint model is able to correctly rewrite the query by leveraging the query context.

\begin{table}[t]
    \caption{Human evaluation results for query rewriting relevancy}
    \label{tab:human_eval}
    \centering
    \begin{tabular}{c|ccc}
    \hline 
    & \textbf{Lose} & \textbf{Tie} & \textbf{Win}\\
    \hline
    Joint vs Separate & $22\%$ & $49\%$ & $29\%$ \\
    Joint vs Rule-based & $29\%$ & $60\%$ & $11\%$ \\
    \hline
    \end{tabular}
    
\end{table}


\subsubsection{Comparison with Baseline Methods}
\label{sec:comparison}
We are struggling to find a state-of-the-art query rewriting method with open-source code to compare with. Eventually, the only reasonable baseline we can compare with is a rule-based method as follows.
\begin{itemize}
    \item Rule-based: A baseline method that has been widely used in our system, one of the largest e-commerce search engines in the world. The method starts from a human-curated synonym phrase dictionary. For a given query, it simply replaces the phrase in the query with its synonym phrase from the dictionary, to generate the rewritten query.
\end{itemize}
We use the following evaluation metrics.
\begin{itemize}
    \item \emph{F1 score} is calculated using the prediction precision (p) and recall (r) rate by the standard equation $2pr/(p+r)$. The queries, including the rewritten query and the original query, are both represented by a set of all its unigrams and bigrams. Then we are able to calculate the prediction precision, as the number of overlapping n-grams divided by the number of n-grams in rewritten query, and the prediction recall, as the number of overlapping n-grams divided by the number of n-grams in source query. This F1 score measures the n-gram similarity between the rewritten query and the original query. A higher F1 score indicates  a more similar rewritten query to the original query.
    \item \emph{Edit distance} stands for the Levenshtein distance~\cite{levenshtein1965levenshtein} between the rewritten query and the original one. Smaller edit distance indicates more similar rewritten query to the original query.
    \item \emph{Cosine similarity} calculates the embedding cosine similarity between the rewritten query and the original query. The embeddings are computed from an embedding retrieval model~\cite{zhang2020towards} in our production. Higher cosine similarity stands for more semantic relevancy in embedding space.
\end{itemize}
To sum up, the F1 score and edit distance measure the lexical similarity between the rewritten query and origin query. The cosine similarity measures the semantic similarity between them. In our query rewriting task, the ultimate goal is to retrieval more relevant items. Thus, we are actually looking for two paradoxical directions in metrics: lexical diverse but semantically relevant rewritten queries.

From the comparison results in Table~\ref{tab:comppare with baseline}, We can make the following observations: 1) the rule-based method achieves the highest F1 score and lowest edit distance, which indicates very high lexical similarity between the rewritten query and the original query. Though the highest cosine similarity indicates the good semantic relevancy, it is still not optimal from the perspective of query rewriting. Since those lexical similar rewritten queries won't contribute much to retrieve more relevant items.
2) The separately trained model and the jointly trained model have similar performance on all metrics, and the latter shows slightly better cosine similarity, which indicates slightly better semantically relevant rewritten queries, though the latter shows much better relevancy by human evaluation. But both models show much more diverse queries are generated, \ie, much lower F1 score and much higher edit distance, while still maintaining good semantic relevancy, \ie, slightly lower cosine similarity, comparing with the rule-based one.

    
    
    

\begin{table}[tb]
    \caption{Comparison between baseline methods with our proposed methods}
    \label{tab:comppare with baseline}
    \centering
    \begin{tabular}{c|ccc}
    \hline 
    & \textbf{F1-score} $\uparrow$ & \textbf{Edit Distance} $\downarrow$ & \textbf{Cosine Similarity} $\uparrow$\\
    \hline
    Rule-based & $0.676$ & $1.767$ & $0.711$ \\
    Separate & $0.193$ & $5.340$ & $0.660$ \\
    Joint & $\mathbf{0.254}$ &  $\mathbf{4.821}$ & $\mathbf{0.668}$ \\
    \hline
    \end{tabular}
    
\end{table}


\subsection{Online Experiments}
\label{sec:online}

We would like to focus on the overall improvement of a search system using the proposed model as an additional retrieval method. We conducted live experiments on $10\%$ of the entire site traffic during a period of 10 days using a standard A/B testing configuration.

In the control setup (baseline), it includes all the candidates available in our current production system, which are retrieved by inverted-index based methods with a standard query rewriting system. In the variation experiment setup, it generates at most $3$ rewritten queries, each of which retrieves at most $1,000$ candidates in addition to those in the baseline. For both settings, all the candidates go through the same ranking component and business logic. The ranking component applies a state-of-the-art deep learning method~\cite{li2019dl}. Here, we emphasize that our production system is a strong baseline to be compared with, as it has been tuned by hundreds of engineers and scientists for years, and has applied state-of-the-art query rewriting and document processing methods to optimize candidate generation.

Table~\ref{tab:abtest} shows the A/B test results of the jointly trained model. To protect confidential business information, only relative improvements are reported. As we can see, the core e-commerce business metrics, including user conversion rate (UCVR), and gross merchandise value (GMV), as well as query rewrite rate (QRR), are all significantly improved. This A/B test results demonstrate the effectiveness of the proposed method. We hopefully will do more analysis to get insights on how the model improves online results, after we have a longer period for A/B test.

\begin{table}[h]
    \caption{10-days online A/B test improvements}
    \centering
    \begin{tabular}{c|r r r r}
    \hline
            & \textbf{UCVR}  & \textbf{GMV}  & \textbf{QRR} \\
    \hline
    Joint  & $+0.5219\%$ & $+1.1054\%$ & $-0.0397\%$ \\
    \hline
    \end{tabular}
    \label{tab:abtest}
\end{table}

\section{Conclusion and Future Work}
\label{sec:discussion}
In this paper, we have proposed a novel deep neural network model to perform query rewriting in an industry scale e-commerce search engine. Specifically, 1) we formulated the long existing query rewriting into a novel cyclic machine translation problem, in order to leverage abundant click log data to train state-of-the-art machine translation models for this task. 2) We improved the bare-bones algorithm of separately training two translation models by introducing a cyclic consistent likelihood, which encourages a given query can ``translate back'' to itself in this cyclic translation process. 3) We proposed a system optimization by merging syntax trees into one, which makes this proposed method feasible in practice, \ie, an industrial e-commerce search system. 4) We demonstrated in the ablation study that the proposed method can effectively generate semantically relevant but more standard queries, especially for vague long tail queries, in offline experiments that the cyclic consistent training could boost the model performance in terms of ``translate back'' accuracy, log probability and perplexity,
and in online experiments that the proposed method could significantly improve all business core metrics in one of the world's largest e-commerce search engine.

Apart from the proposed model, we have also explored another promising approach, by leveraging the pre-trained GPT2 model~\cite{radford2019language}, which is a deep transformer-based language model trained on very large data.
In the query rewriting scenario, we can add a special token between the query and title, \ie, ``query \textless sep1\textgreater \ title \textless sep2\textgreater \  query2'', and treat the whole sequence as a ``special'' language. Thus, the GPT2 model could be fine-tuned to learn the language model for this ``special'' language, which hopefully could generate a synthetic title for a given query, then generate a synthetic query from the title.
This approach looks promising since the GPT2 model could benefit from pretraining on very large data. In practice, we, however, have not found it performs better than our jointly trained machine translation models yet. This is one of our future work, and we look forward to more inspirations in this direction.

Our future work also includes explorations on more novel decoding methods, to generate more diverse and likely sequences. For example, diverse beam search~\cite{vijayakumar2016diverse} is another way to increase diversity by optimizing a diversity-augmented objective directly.


Furthermore, developing more reasonable offline metrics to guide our offline model improvement is another important direction. We have found that neither the lexical similarity (F1 score and edit distance) nor the semantic similarity (cosine similarity) aligns well with the query rewriting objective, which is essential to generate diverse and semantically relevant queries. We believe such a metric could greatly benefit our model development.





\bibliographystyle{IEEEtran}
\balance
\bibliography{bibliography}

\end{document}